\documentclass[12pt,final]{iopart}
\newcommand{\W}{14cm}
\usepackage{iopams}
\usepackage{graphicx}

\begin{document}

\title[Geometrical and transport properties of single fractures]{Geometrical and transport properties of single fractures: influence of the roughness of the fracture walls}

\author{H. Auradou}

\address{Univ Pierre et Marie Curie-Paris6, Univ Paris-Sud, CNRS, F-91405.
   Lab FAST, Bat 502, Campus Univ, Orsay, F-91405, France.} \ead{auradou@fast.u-psud.fr}
\begin{abstract}
This article reviews the main features of the transport
properties of  single fractures. A particular attention paid to
fractures in geological materials which often display a roughness covering a
broad range of length scales. Because of the small distance separating the
fracture walls, the surface roughness is a key parameter
influencing the structure of the void space.
Studies devoted to the characterization of the surface
roughness are presented as well as works aimed at characterizing the
void space geometry. The correlation of the free space is found to be
crucially function of the failure mechanism (brittle, quasi brittle or
plastic...) but also of possible shear displacements during the failure.
The influence of the surface roughness on the mechanical behavior of
fractures under a normal load and a shear stress is also described. Finally,
experimental, numerical and theoretical works devoted to the study of the
influence of the fracture void geometry on the permeability and on the
hydrodynamic dispersion of a dissolved species are discussed.
\end{abstract}

\maketitle
\section{Introduction}
Since fractures are difficult to avoid and may extend in size over scales ranging from microns to the many meters, their flow and transport properties have received a lot of attention. Most of the past effort were focussed on fractured in geological materials because they are of interest in a number of context ranging from oil or water reservoir management, geotechnical applications such as underground storage of wastes and deeper Earth systems such as earthquakes \cite{NAS}. Amongst others applications, we can cite the long term durability of concrete structures or sealing problems in mechanical industry and more
particularly in the design and safety of nuclear power plants or cryotechnic
rocket engines, where the sealing of some units is performed by a direct contact
between metallic rough surfaces \cite{Marie2003,Plouraboue2006}.\\
All the past studies revealed that transport of passive species
and mechanical properties of single fractures depend almost
exclusively on the geometry of their void space \cite{NAS}.
Moreover, because existing fractures provide planes of weakness on which
further deformation can more readily occur, fractures also largely control the mechanical behavior of materials.
Finally, fractures also cause the behavior of large volumes of
material to be different from that of small scale samples tested in laboratory.\\
Past studies clearly point out two important characteristics of flow through fractures:
The first is that even under normal loads as large as $100 MPa$ for rocks \cite{Watanabe2008}
or $600 MPa$ for pressed metallic joint \cite{Marie2003},
the permeability of the fractures is still higher than the matrix porosity.
This demonstrates that the walls of these fracture do not perfectly match so that a sufficient free space inside which flow takes place is left. The second key feature is the strong channelization of the flow along few preferential paths of low hydraulic resistance \cite{Brown1998,Dijk1999}. Even if research efforts were devoted to this subject, the relation between the void space geometrical characteristics and the fracture permeability is still an open question, largely because of the localization of the flow; the latter has also a strong impact on the transport of particles or contaminants. In the latter case, early breakthroughs of contaminants are often reported for flow in fractured materials: the accurate description of such phenomena is still not achieved.\\
In this paper, we report briefly recent developments together with the most important past results concerning the description of the transport properties of single fractures.\\

The first section of this paper deals with the statistical properties of the aperture field of single fractures either made artificially or sampled on field sites. Although the aperture field have a very variable topology, they can be characterized by a correlation length. Its value is found to be crucially related to the genesis of the fracture: For instance, plastic deformation and small scale disaggregation following failure prevent the fracture walls from mating perfectly at small length scales, in this case fractures are characterized by a small - compared to the fracture size - correlation length reflecting past physical processes. Small displacements between the fracture walls also prevent the fracture from mating. We will show that because of the surface roughness that may extend over scales comparable to the fracture size, the aperture is anisotropic with a large correlation length in the direction perpendicular to the shear displacement direction. The evolution of the void space as well as the geometry of the contact zones are reported for various stress conditions (normal load or shear stress).\\
The void space is a key parameter controlling the transport property of fractures. For instance, the amount of water that may flow under a pressure gradient through a fracture is crucial in many applications. Yet, all past experimental observations found that this quantity cannot be simply related to the geometrical parameters such as the mean fracture aperture. This problem as attract a lot of attention but no satisfactory model as been yet found.\\
The heterogeneous nature of the aperture also strongly influences the transport of solute particles: because of preferential channels inside which a large part of the flow takes place, an important fraction of solute is transported by convection over large distances rapidly while the rest of the solute lags behind resides in reduced velocity regions. As a result, classical description of dispersion using a Fickian approach breaks down at the fracture scale. The studies performed on that topic are described in the last section (Sec.~\ref{sec:dispersion}) of this paper.\\
A key parameter controlling the aperture is the roughness of the fracture walls. The following section discusses results concerning the statistical description of the roughness of fracture surfaces. Past results on single fractures propagating in disordered systems such as rocks report that the roughness increases non linearly with the fracture size. We will then show how this multiscale roughness influences the transport properties of fractures.
\section{Roughness of the fracture walls}
\label{sec:roughness}
\begin{figure}
\includegraphics[width=\W]{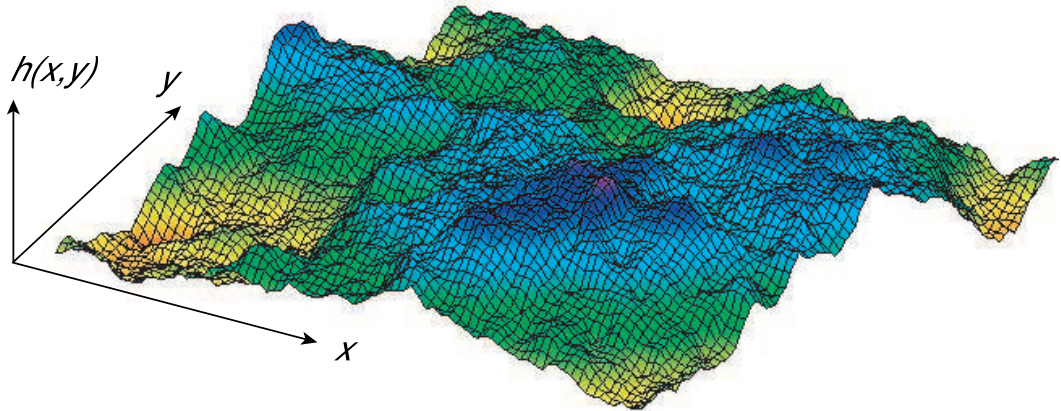}
\caption{Map of a fractured surface of a granite block broken under a tensile stress. The heights $h(x,y)$ were recorded by a mechanical profilometer surface. The size in the $(x,y)$ plane of the surface is $200$ mm $\times 200$ mm, over this area rms of the heights is $3.5$ mm and the height difference between the highest and lowest points is $20$ mm.}
\label{fig:fig1}
\end{figure}
\begin{figure}
\includegraphics[width=\W]{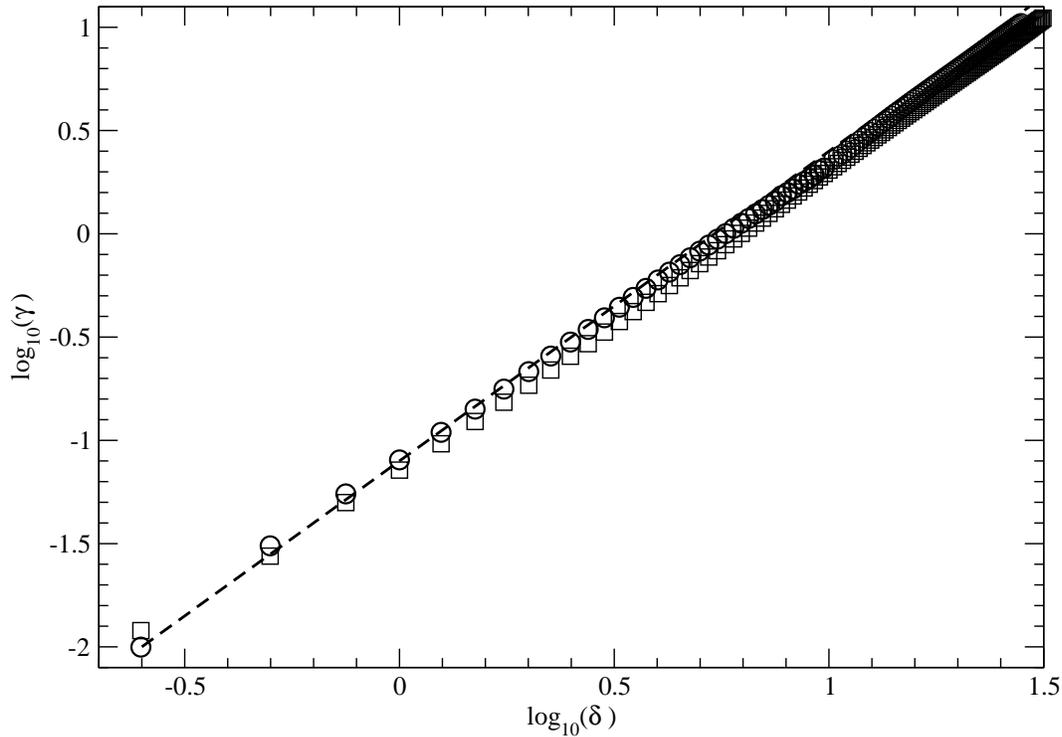}
\caption{Variation of the log of the correlation function $\gamma$ (mm$^2$) as function of the log of the lag distance in $|\boldsymbol{\delta}|$ (mm) for vectors oriented along the direction of propagation of the fracture font: circles and squares are for measurements normal to that direction. The dashed line has a slope of $1.5$. The surface studied is the granite surface of size $200 \times 200$ mm displays in Fig. ~\ref{fig:fig1} \cite{Auradou2001}.}
\label{fig:fig2}
\end{figure}

The structure of the aperture field is strongly related to the relative position of the fracture together with the statistical distribution of the asperities.
Naturally occurring rough surfaces are characterized by a roughness present over a broad range of length scales.
As illustrated by figure \ref{fig:fig1}, these correlations typically manifest themselves through the increase of the amplitude of the roughness with the observation scale considered. Statistical analysis of the scaling dependance of the surface roughness on the fracture size reveals that the roughness of fractured surfaces is well described as a self-affine fractal structure \footnote{Numerous works have characterized the surface roughness through fractal dimensions $D$. The values reported agreed with the inferred values of $\zeta$'s; henceforth no direct reference to $D$ will be made. The difference between this two exponents is discussed in \cite{Mandelbrot1985}.}.
This means that the spatial distribution of the local heights $h(\textbf{r})$ is statistically invariant under the transformation
$\textbf{r} \rightarrow \lambda \textbf{r}$ and $h \rightarrow \lambda^\zeta h$ where $\zeta$, the self-affine or Hurst exponent, ranges from $0$ to $1$.
So, giving the correlation function $\gamma (\boldsymbol{\delta})=\left\langle \Delta h^2 \right\rangle$ to find a height difference $\Delta h=h(\textbf{r})-h(\textbf{r}-\boldsymbol{\delta})$
over an horizontal vector $\delta$, the scaling invariance leads to:
\begin{equation}
\gamma(\lambda \boldsymbol{\delta})=\lambda^{2 \zeta} \gamma(\boldsymbol{\delta}).
\end{equation}
Figure \ref{fig:fig2} shows the variation of the function $|\vec{\gamma}|$ computed for distances $\boldsymbol{\delta}$ oriented along (circles in Fig. \ref{fig:fig2}) and normal (squares in Fig. \ref{fig:fig2}) to the direction of propagation of the fracture front. For both direction, $\gamma$ is a monotonously increasing function of the lag distance $|\vec{\delta}|$ considered indicating a clear increase of the surface roughness with its extension. In addition, the increase is linear indicating that $\gamma$ follows a power law of $\delta$ with an exponent $\zeta=0.75\pm0.05$ which  shows only a weak dependance with the orientation considered. Exponents close to $0.8$ have been reported in a large number of previous studies and the hypothesis of its universality was furthermore conjectured \cite{Bouchaud1990}.\\
More recent works indicate that this exponent may take two different values, depending on the type of fracture materials: $0.8$,
for most of the materials including glass, cements, granite, tuff
\cite{Brown1986,Maloy1992,Poon1992,Schmittbuhl1993,Glover1998,Auradou2005,Matsuki2006} while $\zeta \simeq 0.5$ for materials such as sandstone
\cite{Boffa1998,Ponson2007}, calcite \cite{Gouze2003} or materials with similar microstructure
such as sintered glass beads blocks \cite{Ponson2006}.
The geometry of the microstructure is not the sole parameter controlling the roughness. Other experiments performed using sandstone reported a roughness exponent close to $\zeta=0.75\pm0.03$ \cite{Nasseri2006} disagreing with the observation of \cite{Boffa1998,Ponson2007}. However, in this work \cite{Nasseri2006} a fracture process zone measured by acoustic emission ahead of the fracture front was actually present while in the latter experiments \cite{Boffa1998,Ponson2007} such zone was not existing. The existence of damage zone where brittle micro cracks spread affecting the growing of a single brittle crack  was then found to be a key mechanism explaining the existence of two types of fracture roughness \cite{Nasseri2006}. These non linear and plastic effects were further conjectured to correspond to a Family Vicsek self-affine morphology \cite{Ponson2006b}.\\
The effect of the fracture size on its roughness is a key issue which was studied by few authors: Matsuki {\it et al.} \cite{Matsuki2006} performed measurement over metric scale granite fractures that do not revealed any variation of the exponent $\zeta$ with the fracture size. The same conclusion was drawn from the field measurement performed by Schmittbuhl and co workers \cite{Schmittbuhl1993}. Yet, by combining millimetric scale observations with measurements at the metric scale, Brown and Scholtz \cite{Brown1985} still observed that the rms deviations of the height of the fracture walls increases with their length but, this time, with an exponent decreasing with the size of the fracture.

\section{Dependence of the aperture field on mechanical deformations}
\label{sec:aperture}
\subsection{Aperture field reconstruction}
The topography of a single surface does not tell the full story: the key parameter controlling mechanical and transport properties of the fracture is the correlation between the roughness of the two fracture walls. Different methods have been used to map the fracture void space. In the most popular method, the map is reconstructed by using the maps of the heights of the two individual fracture walls as measured by profilometry  \cite{Brown1986,Vickers1992,Glover1998}. This method usually requires to adjust the halves in order to get the best match estimates by, for example, minimizing the rms of the aperture. The resulting field corresponds to the aperture of the fracture under zero normal stress and with the two walls as close as possible to a perfect contact over their full surface. The need of a reconstruction procedure can be avoided by performing injections of fluorescent epoxy into the void space \cite{Gentier1989,Hakami1996}. In this case, the
aperture is reconstructed through image analysis of sections; for practical reasons the section are large making difficult the fracture reconstruction. Measurements were also done an transparent replicas of natural fractures \cite{Yeo1998,Koyama2006,Bauget2008}. Dyed fluid is then injected into the fracture void and the local light absorption is related through an appropriate calibration law to the local fracture aperture.
Recent development of X-Ray tomography and the availability of very intense beams offers a new and efficient way to measure the void structure of porous and fractured media with a spatial resolution that can be as good as $5\mu m$ on centimeter scale objects \cite{Gouze2003,Johns1993,Keller1998,Bertels2001}. Finally, technical improvements together with the development of new surface profilometry techniques (such as atomic force, scanning electron or confocal microscopes and laser scanner) allows now the acquisition of 3d maps of the halves of the fracture \cite{Watanabe2008,Matsuki2006,Koyama2006,Sharifzadeh2008} with a very good precision.\\
With most of these techniques it is possible to determine the statistical distributions of the apertures as well as their spatial correlations. These distributions are usually well fitted by normal \cite{Matsuki2006,Vickers1992,Hakami1996,Yeo1998} or log-normal \cite{Koyama2006,Keller1998} distributions with mean of the order of few hundred microns and standard deviation of the order of $0.1\mu m$ under zero normal load. Distributions adjusted by log-normal distribution law are usually observed on fractures that have been submitted to a normal stress: in this case, the aperture distributions are shifted towards lower values as the surfaces come in close contact: normal distribution may then with increasing stress and contact be fitted by a log-normal distribution \cite{Watanabe2008,Sharifzadeh2008,Oron1998}. The characterization of the correlation through the adjustment of the semivariogram $\gamma$ of the aperture field reveals that most of the field are correlated over length scale ranging from few hundred of microns up to few millimeter independently of the rock type \cite{Watanabe2008,Matsuki2006,Vickers1992,Brown1986,Glover1998,Yeo1998,Keller1998,
Hakami1996,Sharifzadeh2008,Koyama2006,Bauget2008}.\\

\subsection{Mechanical deformation and aperture variations under normal load}
When a normal stress is applied to a preexisting fracture, its aperture reduced and contact zones appear resulting in a non linear variation of the deformation with stress: the rate of deformation is highest at low stress values which reflects an increase of the fracture stiffness. After repeated cycling these non linearities which reflects changes in the void space geometry are still observed \cite{Pyrac1987,Sharifzadeh2008}: some voids get closed and additional surfaces come in contact.
The details of the void space geometry were studied by Pyrak-Nolte and co-workers \cite{Pyrac1987} by means of an injection of a low melting-point alloy allowing one to obtain a cast of the void space.
Pictures of the two surfaces are then taken and digitized. The two maps obtained in this way show the locations where metal has
adhered on each surface, but do not give the value of its thickness. The geometry of the area in contact was also studied by Gentier \cite{Gentier1986} who placed thin film of plastic sheet between the fracture walls.
When squeezed, the optical quality of the film is locally altered: image analyzing is then used to determine the region where the fracture walls are in contact. Experiments repeated under normal stress up to $80\ MPa$
indicate that contact areas are heterogeneously spread over and represent a total area of the order of $60\%$ of the fracture area \cite{Watanabe2008,Gentier1986,Pyrac1987}. Recently Sharifzadeh {\it et} \cite{Sharifzadeh2008} report for a fractured granite block contact area fractions of the total surface corresponding to the contact area ranging from $86\%$ for a normal stress of $1\ MPa$ up to $99\%$ at $10\ MPa$. The latter value indicated a near perfect match between the two surfaces.\\
\subsection{Mechanical deformation under shear}
 Recent experimental investigations were intended to determine, visualize and interpret aperture distribution under various shear displacements and different normal stress variation histories. All these works demonstrate that the application of a shear stress alter significantly the void space. Two main features were observed: a dilatancy of the fracture resulting in an increase of its aperture and in the appearance of damage zones. The relative magnitude of the two phenomena is controlled by the normal stress. If the latter is low compared to the rock strength than gouge material is not be produced and protuberant asperities behave like rigid material \cite{Sharifzadeh2008}. The opposite situation - normal stress overcoming the strength of the material - is achieved by, for instance, using cement mortar cast of rock fractures \cite{Gentier2000,Koyama2006}.\\
When a shear stress is applied to one of the fracture walls, small variations of the fracture aperture are first observed up to a shear
displacement of a few millimeter \cite{Olsson1993,Esaki1999,Gentier2000,Sharifzadeh2008} because of the motion is stopped when the asperities of the upper and lower surfaces come in contact. This
is followed by an uplift of the two halves reflecting the dilation, as the shear stress increases simultaneously up to a peak. After that, the shear stress rapidly decreases to a residual value and becomes almost constant. At a critical point near the peak value of the shear stress, for which the latter becomes equal to the strength of the asperities in contact, all the asperities in contact are simultaneously sheared. As a result, the aperture distribution suddenly changes because the roughness of the two surfaces do not match any more.
Increasing further the shear results in a further increase of the aperture and in a reduced contact area.
For high strength materials, the fraction of the area where the surfaces are in contact drops abruptly when the peak value of the stress is reached \cite{Sharifzadeh2008,Watanabe2008}. Meanwhile, the mean aperture increases by more than an order of magnitude \cite{Sharifzadeh2008} corresponding to an increase of the permeability by three orders of magnitude \cite{Watanabe2008}. For more brittle materials, damage is initiated in areas of the fracture with the steepest slope and their extensions in the fracture plan are strongly related to the amount of vertical displacement \cite{Gentier2000}. The gouge particles created during the shear displacement may in turn influence the fractures properties \cite{Amitrano2002}.\\
\subsection{Modifications of the aperture field under shear}
\begin{figure}
\includegraphics[width=\W]{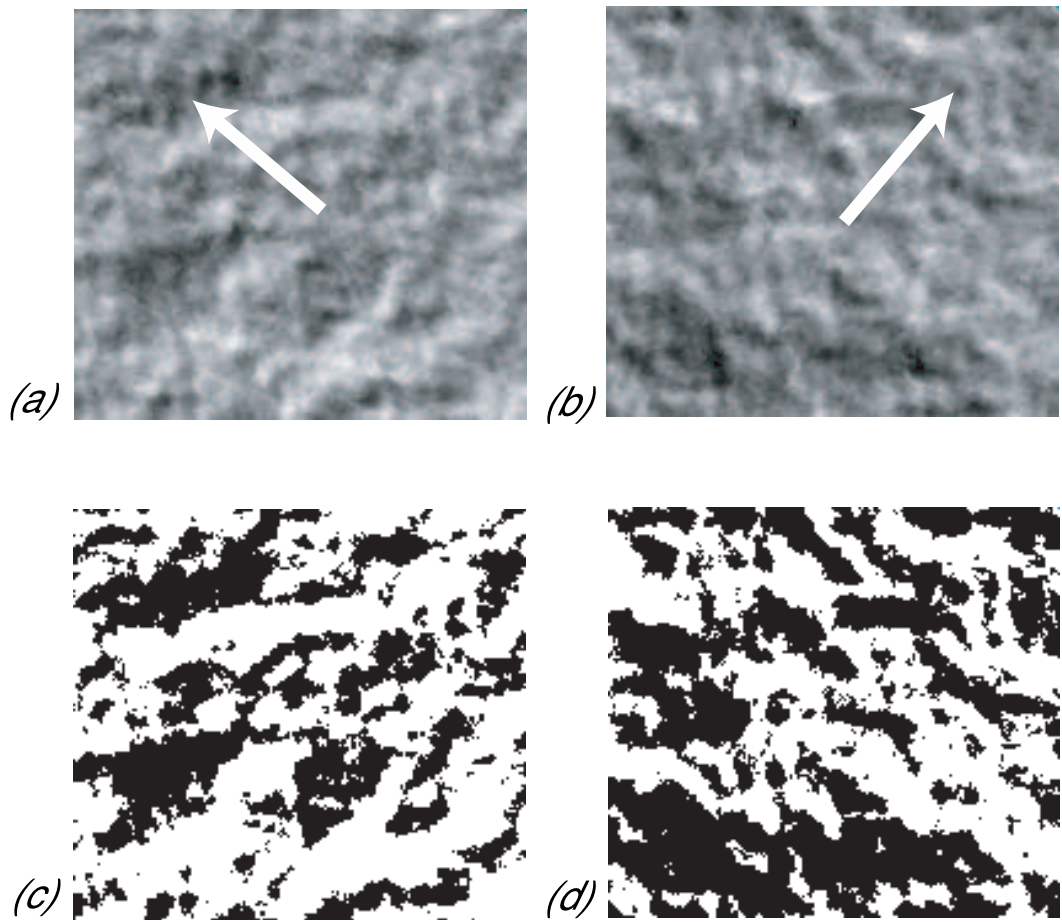}
\caption{Aperture field obtained by displacing a surface height map of a granite fracture according to Eq.~\ref{eq:a}. The magnitude of the shear displacement is in both cases $u/L=0.04$ where $L=60\ mm$ is the fracture size. The shear directions are indicated by arrows. Figures ($c$) and ($d$) represent the aperture for a threshold value set to half the mean fracture aperture (white: $a<\bar{a}/2$, black: $a>\bar{a}/2$).}
\label{fig:fig3}
\end{figure}

\begin{figure}
\includegraphics[width=\W]{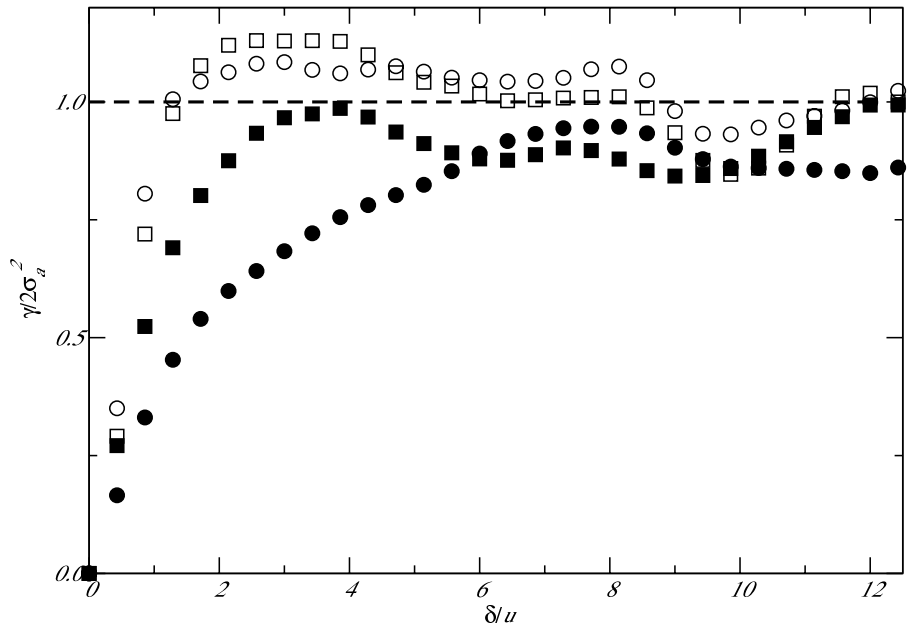}
\caption{Variation of $\gamma$ normalized by $2\sigma_a^2$ where $\sigma_a$ is rms of the aperture as function of the lag $|\vec{\delta}|$ normalized by the value of the shear displacement $u$ for direction perpendicular (filled symbols) and along (open symbols) the shear displacement. The fields are shown in Fig.~\ref{fig:fig3}: circles and squares correspond respectively to field $(a)$ and $(b)$}
\label{fig:fig4}
\end{figure}

When the fracture halves are brought together so they are separated by a minimum free space, the aperture frequency
distribution follows a normal or a log-normal distributions and the regions in contact are evenly distributed. When the shear increases, the
aperture distribution is shifted towards the high aperture values indicating an increase of the mean fracture aperture. In the same time, the distribution broads and becomes well adjusted by a normal law \cite{Watanabe2008,Olsson1993}. Under such conditions, the contact areas are then spatially localized and an heterogeneous distribution of the aperture is observed. Moreover, the analyze of the spatial correlation shows that the aperture field is characterized by a strong anisotropy with a correlation length which is much larger in the direction perpendicular to the displacement \cite{Watanabe2008,Yeo1998}.\\
This crucial shift in the aperture field is illustrated quantitatively by considering the simplified configuration in which fractures made of complementary walls that are separated both vertically in order to open the
fracture and laterally in order to mimic shear displacements (if
any). Under such conditions, if $\bar{a}$ is the mean aperture and
$\vec{u}$ the relative shear displacement  in the plane of the
fracture, the  local aperture at a point $\vec{r}$ satisfies :
\begin{equation}
a(\vec{r}) = a(\vec{r}) - a(\vec{r}+\vec{u}) + \bar{a}. \label{eq:a}
\end{equation}
This equation allows one to determine the aperture field $a({\vec r})$
for a given shift ${\vec u}$ from the experimental surface profile
map $a({\vec r})$ assuming that the surface match perfectly for zero normal and lateral shifts.
The Fig.~\ref{fig:fig3} shows examples of aperture fields computed from the heights map of a granite fracture. The size of the fracture is $60\ mm$ and lateral shifts representing only $4\%$ of the fracture size were applied along the two diagonal of the fractures.
Even if the displacement is small, the aperture fields show a remarkable anisotropy with a strong correlation in the direction perpendicular to the displacement.
Such an anisotropy was also observed by Matsuki and co-workers \cite{Matsuki2006} in the case of two walls which do not match perfectly under zero shear displacement as soon as a shear displacement large enough to generate heterogeneities with a correlation length overcoming the correlation length observed in the absence of displacement was imposed.
The correlations in the aperture field can be
characterized through the correlation function $\gamma(\vec{\delta})$, defined in Sec.~\ref{sec:roughness}, which measures the spatial correlation of the aperture field between two points separated by the 2D vector $\vec{\delta}$. The orientations of the lag $\vec \delta$ respectively parallel and perpendicular to the shear
displacement ${\vec u}$ are of special interest (in the following,
the lags are respectively referred to as $\parallel$ and $\perp$).
Fig.~\ref{fig:fig4} displays the semivariances $\gamma(\delta_\parallel)$ and $\gamma(\delta_\perp)$ of the aperture fields display in Fig.~\ref{fig:fig3}.
For small magnitudes $\delta$ and for a given exponent, the apertures are very similar (high spatial correlation
at short distances) and $\gamma$ is small. As $|\delta|$ increases, so does $\gamma$, because the spatial correlation of the aperture decreases. If $|\delta|$ increases further, up to a value higher than the correlation length of the
aperture field, the apertures become uncorrelated and $\gamma$ should reach a same saturation value for both orientations of $\delta$ equal to twice the aperture variance $\sigma_a^2=<(a(x,y)-\bar{a})^2>_{x,y}$.
This value corresponds to a ratio $\gamma/2 \sigma_a^2=1$ (dashed line in Fig.~\ref{fig:fig4}) which is actually the asymptotic value observed at high $|\vec{\delta}|$ values on Fig.~\ref{fig:fig4}.
Even though the asymptotic values are the same, the way in which they reach this saturation value differs greatly depending on the orientation of $\vec{\delta}$ with respect to the shift.
The normalized semivariances corresponding to the perpendicular direction, never exceed the saturation value and reach it in an overdamped way. On the other hand, the functions $\gamma/2\sigma_a^2$ for $|\vec{\delta}|$ parallel to the shear displacement display a small overshoot, reflecting a local anticorrelation. These different behaviors result directly from the large-scale anisotropic structures observed in Fig.~\ref{fig:fig3}.\\

The history of the stress variations experienced by the fracture has a strong influence on its aperture. Under high normal load, the contact area between the two halves is important and the contact zones are distributed heterogeneously in the fracture plane. As a consequence the flow and the geometry of the path of least resistance will be greatly influenced by the fracture spatial distribution of the void space. When the fracture walls have undergone, in addition, a shear displacement, high aperture channels develop in the direction normal to the shear and represent paths of reduced hydraulic resistance. The following section reviews some of the studies and the most important findings concerning the flow of a single liquid in a fracture.

\section{Experimental and numerical estimation of the fracture transmissivity}
As a first approximation, it is natural to view
fractures as made of parallel flat plates with a separation equal to the average mean distance separating the fracture walls. In this case, the macroscopic transmissivity is proportional to the cube of this aperture. Yet, a large number of experimental measurements of the transmissivity of
fractured rocks reveals that the transmissivity decreases more
rapidly than the cube of the mean aperture \cite{NAS}. These findings
demonstrate that as soon as the fractured surfaces are in contact (or close to contact)
the complex geometry of the structure of the free space where flow
takes place strongly alter the flow, leading to an increase of the
hydraulic resistance. This non linear relation between the cube of the fracture aperture and
the transmissivity results from the strong localization
of the flow along a few preferential paths which is found
to increase with the degree of heterogeneity.
A classical way to study the coupling between the fracture geometry
and the flow structure is through the Reynolds approximation which
- by neglecting the out of plane flow movement - reduces the $3D$ flow problem into a $2D$ problem.
The next section summarizes the hypothesis
used to derive the Reynolds equation from the Navier-Stokes in the context of flow in fractures.\\
\subsection{Equation of motion of the fluid motions and definition of the fracture transmissivity}
Three elements are necessary to describe the motion of a fluid: the
conservation of mass and momentum and the boundary conditions. For
an incompressible fluid (this is the case for most of the liquids),
the mass conservation equation reduces to $\bi{\nabla}.\bi{v}=0$ where
$\bi{v}$ is the velocity vector describing the motion of the fluid
particles. For a Newtonian fluid, the
momentum equation reduces to the Navier-Stokes equations given
by \cite{Hulin}:
\begin{equation}
\rho \frac{\partial \bi{v}}{\partial t}+(\bi{v}.\bi{\nabla})\bi{v}=-\nabla p +\mu \nabla^2\bi{v},
\end{equation}
where $\mu$, $\rho$ and $p$ are respectively the dynamic viscosity of the fluid, its density and the total pressure.\\
In most flows through fractured media, the effect of inertia can be
considered as negligible to that of the viscosity. This can be made
quantitative through the estimation of the Reynolds number $Re$
 given by:
\begin{equation}
Re=\frac{\rho U \bar{a}^2}{\mu l}
\end{equation}
where $l$ is a characteristic distance over which the aperture varies
noticeably in the direction of the flow \cite{Zimmerman1991}. When
the inertia term is neglected ($Re<<1$) and in the steady state
regime, the Navier-Stokes equations can be reduced to the so called Stokes
equations \cite{Hulin}:
\begin{equation}
-\bi{\nabla} p + \mu \nabla^2 \bi{v}=\bi{0}.
\label{eq:stokes}
\end{equation}
This assumption breaks down in locations where the apertures displays fast variations: then, the effect of the inertial must be taken into account and, in regions where the velocity decreases abruptly, recirculation eddies appear. These effects result in an additional dissipation which influences in turn the velocity field, leading to the reduced permeability observed experimentally \cite{Dijk1999,Cardenas2007,Yiguang2008,Skjetne1999}.
However, recent experimental observations \cite{Zimmerman2004} demonstrated that such non linear flow regimes are observed at the fracture scale only for Reynolds number greater $10$.\\
Together with the mass conservation equation, the three components of the Stokes equation form a set of $4$ coupled equations and that can
be solved analytically only for very simple cases. The most widely used hypothesis reducing
this $3D$ problem to a $2D$ one is to
assume that the tortuosity of the flow lines across the local aperture
({\it{i.e.}} along $z$ in Fig.\ref{fig:fig1}) is negligible. This is equivalent to consider that the pressure variation across the gap is small in regard to the pressure gradient in the fracture plane {\it{i.e.}} $\frac{\partial p}{\partial z}<<\frac{\partial p}{\partial y}$ (lubrication approximation). As a result, only the components of the velocity vector aligned with the pressure gradient are non zero. Together with a dimensional analysis of the variation of the velocity with respect to the direction \cite{Zimmerman1991}, one show that Eq.(\ref{eq:stokes}) reduces then to:
\begin{equation}
\bi{\nabla p(x,y)} = \mu \frac{\partial^2 \bi{v}}{\partial z^2}.
\end{equation}
Integrating this equation with respect to $z$ bearing in mind the no slip boundary
conditions at the fracture walls yields to a velocity profile which is parabolic
(the velocity being null at the walls and maximum half way between them).\\
The integration of the velocity profile in the fracture gap leads to a estimation of the flow rates per unit of fracture width $Q_x$ and $Q_y$ which are given by:
\begin{eqnarray}
Q_x=- T(x,y) \frac{\partial p}{\partial x}\\
Q_y=- T(x,y) \frac{\partial p}{\partial y},
\label{eq:trans2}
\end{eqnarray}
where $T(x,y)=\frac{a(x,y)^3}{12 \mu}$ is the local fracture transmissivity in $m^2.s^{-1}$. Together with the mass conservation, $\bi{\nabla}.\bi{Q}$, the previous set of equations reduce to the Reynolds equation:
\begin{eqnarray}
\bi{\nabla}.(T(x,y) \bi{\nabla}p)=0
\label{eq:Reynolds}
\end{eqnarray}
where the derivatives are taken over the $(x,y)$ plane. \\
\subsection{Transmissivity of natural fracture and its modeling}
Quantitative comparisons between experimental measurements and numerical estimations were rarely performed in configurations in which the detailed of the geometry of the void space was known. Most of these works obtained the macroscopic transmissivity by solving the Reynolds approximation (See Eq.(\ref{eq:Reynolds})). This approximation, even if it usually leads
to an overestimation of the flow velocity or equivalently of the hydraulic aperture that increases when the fracture aperture is reduced \cite{Yeo1998,Bauget2008,Konzuk2004,Meheust2000}, is used extensively to deal with flow problems in fractures \cite{Glover1998}. Mourzenko {\it{et al}} \cite{Mourzenko1995} reported systematic comparisons between numerical solutions for flow in rough fractures obtained using the Reynolds approximation (see Eq.(\ref{eq:Reynolds})) and the 3D Stokes equations given by Eq.(\ref{eq:stokes}). The simulations are found to give similar results only for wide and smooth fractures; but, when the fracture aperture is decreased or the aperture fluctuation is increased important differences between the two sets of results are observed. Comparable conclusions were obtained by Brown {\it{et al}} \cite{Brown1995} when comparing the two types of solutions for flow in sinusoidal channels. The geometrical validity of the application of the Reynolds approximations is discussed in \cite{Oron1998}.\\

Insofar flow through the fracture is modeled in the framework of the the lubrication approximation, the problem of finding the macroscopic transmissivity of the fracture reduces to the problem of finding the effective conductivity of an heterogenous $2D$ conductivity field. Many theories have been developed in the past years to estimate this quantity. Among the most popular is the small perturbation analysis \cite{Gelhar1986}. In the context of fracture flow, this approach allows one to estimate the effective transmissivity for any aperture distribution - including normal, lognormal distribution \cite{Inoue2003}, unidirectional ripples \cite{Zimmerman1991} - by means of a first order expansion given by \cite{Inoue2003}:
\begin{equation}
T = \frac{\bar{a}^3}{12\mu} (1+3 (1-3\alpha) S^2+...)
\label{eq:Transmissivity}
\end{equation}
where $\bar{a}$ is the mean aperture, $S=\sigma_a/\bar{a}$ is the reduce aperture and $\alpha=\int \frac{d\bi{u}}{2\pi^2} \theta(\bi{u})$ where $\theta(u)$ is the Fourier transform of the correlation function $1-\gamma(\bi{\delta})/2\sigma_a^2$ defined in Sec.~\ref{sec:aperture}. Equation (\ref{eq:Transmissivity}) clearly demonstrates that while the magnitude of the macroscopic transmissivity is controlled by the large aperture zones that control $\bar{a}$, the less open zones also play an important part by reducing the transmissivity of the fracture when it is closed.\\
The prediction of Eq.(\ref{eq:Transmissivity}) compare reasonably well with several numerical simulations \cite{Brown1989,Brown1987} and experimental results \cite{Matsuki1999}. Yet, this perturbation theory ignores the possible channelization of the flow \cite{Brown1998,Moreno1988,Tsang1989,Bruderer2001}
, so that deviation from this theory with experimental observation are expected when the heterogeneity of the aperture is enhanced, for instance, when large normal load are applied.\\

Fluid flow through fracture is also hindered by the presence of contact areas which, under a large normal load represent an important fraction of the total fracture area. For fraction lower than  $25\%$ and under the assumption that the fracture aperture is constant outside the contact zone so that the governing equation for fluid flow is a Laplace equation, Zimmermann and co workers \cite{Zimmerman1992} obtained an analytical expression of the effective transmissivity for various simplified shapes of the contact area (circular or elliptical). For contact area occupying a large fraction of the surface, flow is controlled to a large extent by the location and distribution of the constrictions connecting the larger voids into continuous paths \cite{Nolte1988}. It is them tempting to map the void structure of tightly mated fractures walls onto a discrete network. Recently Plourabou{\'e} {\it et al} \cite{Plouraboue2006} developed an efficient method based on mapping the continuous transport equation onto a discrete network of conductance function of the local geometry of the fracture aperture \cite{Plouraboue2004}.\\
\subsection{Transmissivity variations induced by a shear displacement}
As discussed in Sec.~\ref{sec:aperture}, shear displacement causes the  fracture to dilate resulting in an increase of the fracture transmissivity \cite{Olsson1993}. The displacement also induces an anisotropy of the aperture field. The correlation length of the aperture is indeed much larger in the direction normal to the shear displacement then parallel to it. As a result, the transmissivity is also anisotropic with a larger value in the direction perpendicular to the shear. More precisely, numerical \cite{Watanabe2008,Matsuki2006,Drazer2004} and experimental \cite{Watanabe2008,Matsuki2006,Auradou2005,Gentier1997} investigations report an increase of the anisotropy with the shear displacement with a transmissivity increases in the direction normal to $\vec{u}$ while it drops off parallel to $\vec{u}$. When the fracture aperture is large compare to the aperture fluctuation ({\it{i.e.}} small $\sigma_a/a$), Auradou and coworkers \cite{Auradou2005} found that the behavior is adequately reproduced by considering the fracture as made of parallel channel along with the aperture is constant. For this model, in the direction where the aperture constant the transmissivity is proportional to the arithmetic mean of the cube of the aperture ({\it{i.e.}} $T\propto <a^3>$) \cite{Zimmerman1991} while in the direction where the aperture is varies the transmissivity is proportional to the harmonic mean of the cube of the aperture  ({\it{i.e.}} $T\propto <a^{-3}>^{-1}$) \cite{Zimmerman1991}.
Shear displacement also lead to channelization flow: but compared to the previous situation, the channelization is due here to the structural heterogeneity of the aperture rather than to the broad distribution of the local hydraulic transmissivities observed when the fracture get closed.

\section{Hydrodynamic dispersion of dissolved species}
\label{sec:dispersion}
\begin{figure}
\includegraphics[width=\W]{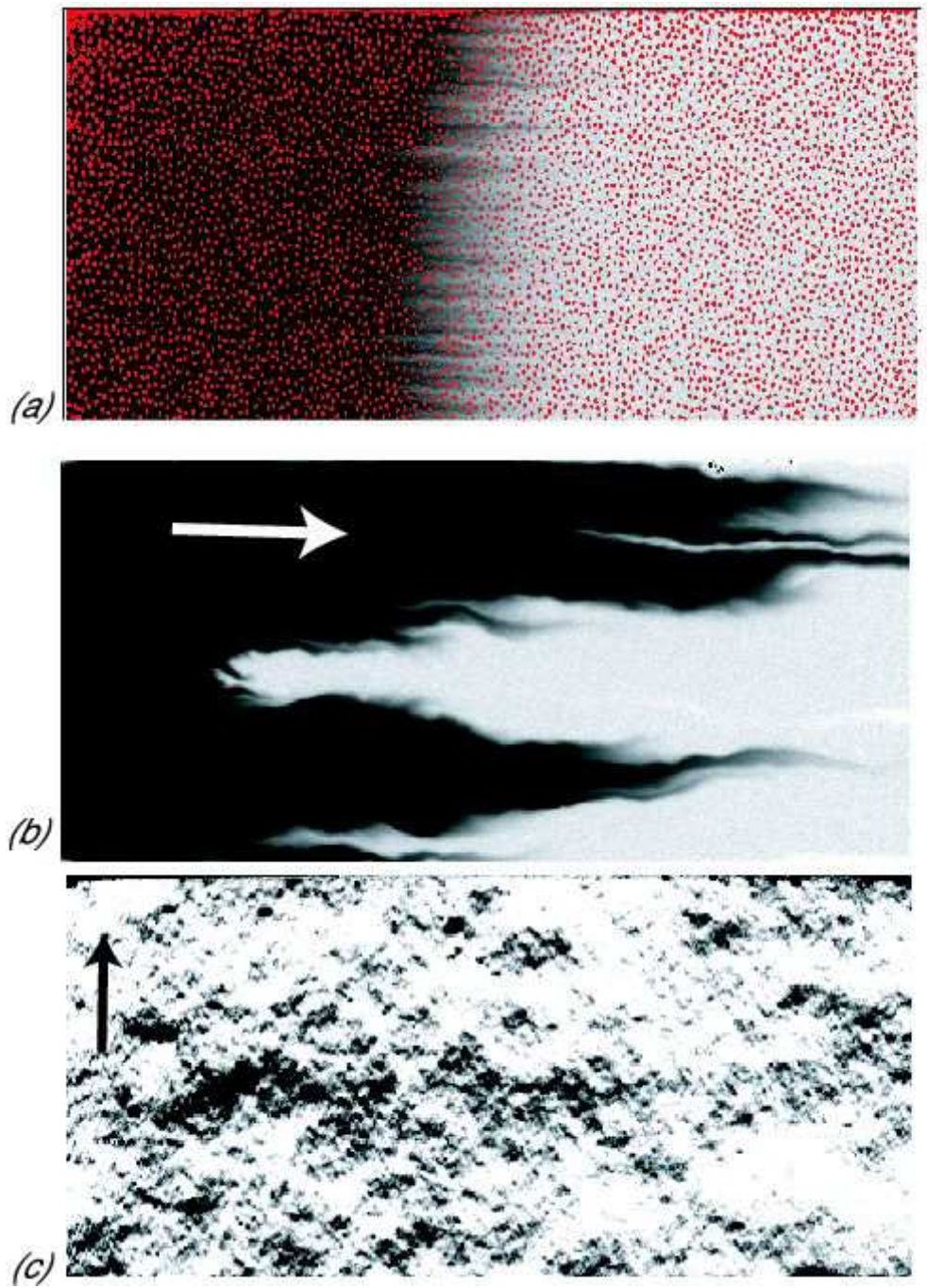}
\caption{Examples of concentration fields coded in gray level observed on two fractures. $a)$ fracture made of a flat plate place above a plate where the roughness is obtained by a random distribution of cylindrical obstacles - dots - (See \cite{Boschan2008} for details). The free space above the obstacle is $0.37\ mm$ and the mean fracture aperture is $0.65\ mm$. $b)$ concentration field when flow occurs in a fracture with sheared walls along the channelized direction. The arrow indicates the flow direction (See \cite{Boschan2007} for details). $c)$ Aperture field of the fracture with sheared complementary walls. Mean aperture: $\bar{a}=0.75\ mm$, and shear displacement $u=0.75\ mm$ (see the black arrow). The shear is perpendicular to the flow direction. White: regions where $a(x,y)$ overcome $\bar{a}/2$ (Black is for $a<\bar{a}/2$.) }
\label{fig:fig5}
\end{figure}

Solute transport is of great interest for instance in fractured rocks because of its implication in
groundwater pollution, $CO_2$ or nuclear waste sequestration in geological formations, and oil recovery but also in corrosion processes where the transport of dissolved species partially controls the leaching rate.
At the scale of a single fracture, dispersion of the dissolved species results from the combined action of the complex velocity field (varying both
in the gap of the fracture and in its plane) and of mixing
by molecular diffusion. The latter allows solute particles
to move from one streamline to another and homogenizes
the spatial distribution of the tracers. In the classical
approach, tracer particles are assumed to perform a
random walk superimposed over a drift velocity.
This latter is the average of the fluid velocity over an appropriate
volume (the representative elementary volume
or REV) while smaller scale variations induce tracer
spreading. At the REV scale, the average $C(x,t)$ of the
tracer concentration over a section of the medium normal
to the mean displacement satisfies the convection diffusion
equation (ADE):
\begin{equation}\label{eq:condiff}
\frac{\partial {C(x,t)}}{\partial t} = U\frac{\partial
{C(x,t)}}{\partial x} +D \frac{\partial^2{
C(x,t)}}{{\partial x}^2}
\end{equation}
where $D$ is the longitudinal dispersion coefficient, $U$ the
mean velocity of the fluid (parallel to x). The value of
$D$ (or equivalently the dispersivity $\alpha = D/U$) is independent
of both time and the traveled distance: it is
determined by the combined contributions of molecular
diffusion (characterized by a coefficient $D_m$) and advection. The relative order of magnitude
of these two effects is characterized by the P\'eclet
number : $Pe = U{\bar{a}}/D_m$.\\
Experiments on cores containing a single fracture \cite{Neretnieks1982,Park1997,Keller1999,Lee2003} focused on the breakthrough curves measured at the outlet and analyzing the dispersion from the overall fracture. The dispersivities determined on the basis of the traditional Fickian advection dispersion models given by Eq.~(\ref{eq:condiff}) were found to be close to values inferred from the small perturbation theory \cite{Gelhar1986,Roux1998}.
The observations suggested that
dispersion is controlled (like in $3D$ porous media \cite{Bear})
by spreading due to velocity variations associated to the
geometry of the void structure. The latter determines the
correlation length of the velocity field, leading to the so
called geometrical dispersion regime.\\
Experiments on transparent cell with walls with a roughness characterized by a small correlation length ($\sim 1\ mm$ compared to pluri centimetric fractures) were them developed in order to study the evolution of the mixing front with the distance \cite{Ippolito1993,Detwiler2000,Boschan2008}. An example of the concentration fields observed in such fractures is shown in Fig.\ref{fig:fig5}a. The influence of small scale heterogeneities located in the mixing zone (central part of the figure) is clearly visible in the concentration field.  These elongate streaks result from the velocity fluctuations created by the obstacles; they have however of small lateral extension so that molecular diffusion is sufficient to smear out these lateral concentrations gradients. The quantitative analysis of the concentration front as of its evolution with time confirms that the dispersion process is characterized by a dispersivity which does not evolve with time (expect close to the fracture inlet). Furthermore at low P{\'e}clet numbers but larger than 1 (so that pure
longitudinal molecular diffusion is negligible), the dispersivity is found to be constant and close to the value inferred from perturbation analysis \cite{Boschan2008}. However, as the P{\'e}clet becomes larger a second regime in which the dispersivity increases linearly with $Pe$ is observed. This new regime corresponds to the Taylor dispersion mechanism \cite{Roux1998} arising from the velocity profile in the fracture gap (the velocity is maximum half way between the walls and cancels out at the boundaries.) that creates concentration gradient within the gap which is reduced through molecular diffusion \cite{Aris1956}.\\
However, these experiments conduct in transparent models with a short correlation length of the fluctuations of the aperture were not able to reproduce the steps observed on breakthrough curves \cite{Neretnieks1982} due to localization of most of the flow in a few preferential channels. Recently
experiments on transparent replica of a real fractures have been developed \cite{Boschan2007,Bauget2008,Boschan2009} in order to study the evolution of the tracer concentration as function of the traveled distance in a geometry closer to that of natural fractured rocks.
For sandstone fractures Bauget and Fourar \cite{Bauget2008} found that the solution derived from the ADE equation appears to be unable to model long-time tailing behavior because of the occurrence of flow channels extending along the full length of the fracture. On the basis of such observations, models considering that dispersion mainly results from the velocity contrasts between such macroscopic channels has been developed \cite{Neretnieks1982,Brown1998,Auradou2006,Bauget2008}. In this model the fracture is represented by a stratified medium and the transport between the inlet and the position of observation is characterized through an heterogeneity factor defined as the ratio of the standard deviation of the distribution of the local permeability and of the average permeability of the equivalent medium. Several different distributions of the permeability assigned to each strata were considered log-normal \cite{Neretnieks1982}, normal \cite{Bauget2008} or computed from the aperture field \cite{Auradou2006}.\\
In the study performed by Bauget and Fourar \cite{Bauget2008}, the heterogeneity factor is a fitting parameter obtained by adjusting the breakthrough curves derived from the model with the experimental observations. This factor is found to decrease with the distance indicating a decrease of the velocity contrast between the flow lines corresponding to an increase of interconnections between the flow paths. This study indicate that tracer transport in a single fracture over short distance may be considered as resulting from poorly connected parallel channels but as the distance increase the number of possible interconnection between these isolated channels increases which slowly reduces the influence of the localization of the flow.\\
The situation were the fracture walls are laterally displaced is analyzed in Refs \cite{Boschan2007,Boschan2009}. The corresponding model fractures are characterized by a strong structural heterogeneity with channels of least hydraulic resistance developing in the direction perpendicular to the shear displacement. When flow occurs along that particular direction, solute particles rapidly spread (see the long structures in Fig.\ref{fig:fig5}b) along these channels (white regions in Fig.\ref{fig:fig5}c). A systematic comparison between the geometry of the iso-concentration front $c(x,y,t)=0.5$ and the stratified model \cite{Boschan2007,Boschan2009} demonstrate, for this flow configuration, that the heterogeneity factor constant along the full length of the fracture remains over the entire fracture length. These results are not valid any more when the flow is parallel to the direction of the shear displacement ({\it{i.e.}} perpendicular to the channels) \cite{Boschan2009}: in this case the breakthrough curves are fitted with the solution of the ADE equations and mainly reflect the dispersion induced by the velocity profile in the fracture gap (Taylor dispersion \cite{Aris1956}).\\

In summary, laboratory scale experiments,
as well as theoretical investigations, demonstrate
that the channelization of the flow has a large influence on transport processes. However, the importance of this phenomena is related the structure of the aperture - a fracture characterized by a broad distribution of aperture will more likely display a strongly channelized structure - but also to the flow direction in the context of a fracture characterized by a structural heterogeneity. The channelization creates concentration gradients transverse to the flow, molecular diffusion tends to reduce such inhomogeneities. Thus, if the distance between the channels and/or the P{\'e}clet number are sufficiently small, transport of solute in the direction transverse to the flow direction may be sufficient to reduce the spreading du to the channels: a Fickian dispersion regime is then expected to be observed.

\section{Discussion and conclusion}
This paper has discussed recent works dealing with flow and transport in single rough fractures. In this context, the statistical properties of the fracture aperture is a key parameter. However,
fracture aperture is a difficult parameter to quantify and measure. The difficulty of these measurements is related to the high precision required in order to model the flow field: for instance the permeability of a fracture varies like $a^3$ where $a$ is the fracture aperture.
Previous investigations have shown that aperture is dependent upon stress history, normal displacement,
shear displacement and scale of study. Under zero shear displacement, the roughness of the two fracture walls may not match any more at small scale because of irreversible deformation occurring during the failure process or induced by physical interactions with the host fluid. In these cases   the aperture is randomly distributed and is well described by a normal or a log normal distribution. Under normal load, the fracture closes and the areas in contact are heterogeneously spread over the fracture plane.
During shear, the aperture sharply increased at first - fracture dilation - with the shear displacement until a shear peak is reached. The aperture distribution then becomes spatially localized. The joint surfaces roughness - which is often suitably described by a self-affine multiscale geometry - is the main origin of the spatial localization of aperture distribution. For high strength materials, the contact ratio decreases rapidly and reaches a small value for large shear displacements. For relative brittle materials, asperities of the surface walls are damaged during the shear displacement. The damage zones are created in the region where the fracture walls have a steep slope creating. The surface roughness is thus - again - a key parameter controlling the spatial repartition of these zones. The latter are heterogeneously distributed over the surface.\\

The flow of the fluid in a fracture is characterized by a strong localization: the flow is concentrated over a small fraction of the fracture area. Channelization arise from two mechanisms: the first result from the widely distributed local hydraulic properties while the second comes from the structural heterogeneity induces by possible shear displacement of the fracture walls.
The channeling phenomenon has important consequence: for instance it imply that solutes transported
in channels have potentially far less
exposure to the rock matrix than in a constant aperture
fracture \cite{Bodin2003}, limiting the reaction, adsorption,
and diffusion of solutes within the rock matrix. In the context of nuclear waste management, the ability of the rock matrix to retard solute movement
through a fractured, crystalline rock may be more limited
than if all the fracture surface was equally available to water
flow. A quantitative estimation of the degree of channelization is still an open issue, but as point out by past studies, their estimation require information on the past hydromechanical history of the fracture.\\

Most of the studies reported in the present paper considered small scale measurements, a few tens of centimeters in length. Detailed measurements on larger scale fractures are mostly limited to measurements of fracture traces \cite{NAS}, they indicate the existence of jumps and discontinuities that come, for instance, from changes in the loading conditions or merging of fractures. These events create longer length scale correlations. Thus at field scales, fracture walls geometry might not appear self-affine and the problem of scale effect is still an issue that needs to be addressed in order to gain an improved understanding of in-situ transport properties.

\ack
The author would like to thank A. Boschan, B. Dido, A. Hansen, J-P Hulin, I. Ippolito and L. Talon for fruitful discussions.
Research efforts of HA are sponsored by the EC through the STREP EGS PILOT PLANT (EC Contract SES-CT-2003-502706) and by the GdR $2990$. This
research was also supported by a CNRS-CONICET Collaborative
Research Grant (PICS CNRS 2178), by the ECOS A03-E02 program and by the I029 UBACyT programs.


\section*{References}
\begin{thebibliography}{10}
\bibitem{NAS}Committee on Fracture Characterization and Fluid Flow, National Research Council 1996 {\it{Rock Fractures and Fluid Flow: Contemporary Understanding and Applications}} ed. The National Academies Press (Washington D.C.)
\bibitem{Marie2003}Marie C, Lasseux D, Zahouani H, Sainsot P 2003 \textit{European Journal Mech. and Env. Eng.} {\bf{48}} (2) 81--86
\bibitem{Plouraboue2006}Plourabou{\'e} F, Flukiger F, Prat M and Crispel P 2006 \textit{Phys. Rev. E} {\bf{73}} 036305
\bibitem{Watanabe2008}Watanabe N, Hirano N and Tsuchiya N 2008 \textit{Water Resour. Res.} {\bf{44}} W06412
\bibitem{Brown1998}Brown S, Caprihan A and Hardy R 1998 \textit{J. Geophys. Res.} {\bf{103}} B3 5125--32
\bibitem{Dijk1999}Dijk P and Berkowitz B 1999 \textit{Water Resour. Res.} {\bf{35}}
3955--59

\bibitem{Mandelbrot1985}Mandelbrot B 1985 \textit{Physica Scripta.} {\bf{32}} 257--60
\bibitem{Bouchaud1990} Bouchaud E, Lapasset G and Plan{\`e}s J 1990 \textit{Europhys. Lett.} {\bf{13}} 73
\bibitem{Brown1986}Brown S, Kranz R and Bonner B 1986 \textit{Geophys. Res. Lett.} {\bf{13}} 13 1430--33
\bibitem{Brown1987}Brown S 1987 \textit{J. Geophys. Res.} {\bf{92}} 1337--47
\bibitem{Maloy1992} Mal{\o}y KJ, Hansen A, Hinrichsen E and Roux S 1992 \textit{Phys. Rev. Lett.} {\bf{68}} 213--15
\bibitem{Poon1992}Poon C, Sayles R and Jones T 1992 \textit{J. Phys. D: Appl. Phys.}
{\bf{25}} 1269--75
\bibitem{Schmittbuhl1993} Schmittbuhl J, Gentier S and Roux S 1993 \textit{Geophys. Res. Lett.}
{\bf{20}} (8) 639--41
\bibitem{Glover1998}Glover P{\it{et al}} 1998 \textit{J. Geophys. Res.} {\bf{103}} B5 9621--35
\bibitem{Auradou2005}Auradou H, Drazer G, Hulin JP and Koplik J
2005 \textit{Water Resour. Res.} {\bf{41}} W09423
\bibitem{Matsuki2006} Matsuki K {\it{et al}} 2006 \textit{Int. J. Rock Mech. Min. Sci.} {\bf{43}} 726--55
\bibitem{Boffa1998} Boffa JM, Allain C and Hulin JP \textit{Eur. Phys. J. Appl. Phys.} 1998  {\bf{2}} 281--89
\bibitem{Ponson2007}Ponson L, Auradou H, Pessel M, Lazarus V and Hulin JP 2007 \textit{Phys. Rev. E} {\bf{76}} (3) 036108
\bibitem{Gouze2003}Gouze P, Noiriel C, Bruderer C, Loggia D and Leprovost R 2003 \textit{Geophys. Res. Lett.} {\bf{30}} (5) 1267
\bibitem{Ponson2006} Ponson L, Auradou H, Vi{\'e} P and Hulin JP 2006 \textit{Phys Rev. Lett.} {\bf{97}} 125501
\bibitem{Nasseri2006}Nasseri M, Mohanty B and Young R 2006 \textit{Pure appl. geophys.} {\bf{163}} 917--45
\bibitem{Ponson2006b}Ponson L, Bonamy D and Bouchaud E 2006 \textit{Phys. Rev. Lett.} {\bf{96}} 3 035506
\bibitem{Brown1985}Brown S and Scholtz C 1985 \textit{J. Geophys. Res.} {\bf{90}} 12,575--12,582
\bibitem{Vickers1992} Vickers B, Neuman S, Sully M and Evans D 1992 \textit{Geophys. Res. Lett.} {\bf{19}} (10) 1029--32
\bibitem{Gentier1989}Gentier S, Billaux D, van Vliet L 1989 \textit{Int. J. Rock Mech. Rock Engng.} {\bf{22}} 149–-57
\bibitem{Hakami1996}Hakami E and Larsson E 1996 \textit{Int. J. Rock Mech. Min. Sci. \& Geomech. Abstr.} {\bf{33}} (4) 395--404
\bibitem{Yeo1998}Yeo IW, De Freitas MH and Zimmerman RW 1998
\textit{Int J Rock Mech Min Sci} {\bf{35}} 8 1051--70
\bibitem{Koyama2006}Koyama T, Fardin N, Jing L and Stephansson O 2006 \textit{Int. J. Rock Mech. \& Min. Sci.} {\bf{43}} 89–-106
\bibitem{Bauget2008}Bauget F and Fourar M 2008 \textit{Journal of Contaminant Hydrology} {\bf{100}} 137--48
\bibitem{Johns1993}Johns R, Steude J, Castenier L and Roberts P 1993 \textit{J. Geophys. Res.} {\bf{98}} (B2) 1889--990
\bibitem{Keller1998}Keller A 1998 \textit{Int. J. Rock Mech. \& Min. Sci.} {\bf{35}} (8) 1037--50
\bibitem{Bertels2001}Bertels S, Dicarlo D and Blunt M 2001 \textit{Water Resour. Res.} {\bf{37}} (3) 649--62
\bibitem{Sharifzadeh2008}Sharifzadeh M, Mitani Y and Esaki T 2008 \textit{Rock Mech. Rock Engng.} {\bf{41}} (2) 299–323
\bibitem{Oron1998}Oron A and Berkowitz B 1998 \textit{Water Resour. Res.} {\bf{34}} (11) 2811--25
\bibitem{Pyrac1987}Pyrak-Nolte L, Myer L, Cook N and Witherspoon P 1987
{\it Sixth Int. Cong. Rock Mech.}
p~ 225--32, ed. Balkema Rotterdam.
\bibitem{Gentier1986}Gentier S 1987 {\it Morphologie et comportement hydromécanique d'une fracture naturelle dans un granite sous contrainte normale, étude expérimentale et théorique}, Thèse de l'Université d'Orléans, 637~p, Edition du BRGM
\bibitem{Gentier2000}Gentier S, Riss J, Archambault G, Flamand R and Hopkins D 2000 \textit{Int. J. Rock Mech. Min. Sci.} {\bf{37}} 161--74
\bibitem{Amitrano2002}Amitrano D and Schmittbuhl J 2002 \textit{J. Geophys. Res.} {\bf{107}} 2375--91
\bibitem{Esaki1999}Esaki T, Du S, Mitani Y, Ikusada K and Jing L 1999 \textit{Int. J. Rock Mech.} {\bf{36}} 641--50
\bibitem{Olsson1993}Olsson W and Brown S 1993 \textit{Int. J. Rock Mech. \& Geomech. Abstr.} {\bf{30}} 845--51
\bibitem{Hulin}Guyon E, Hulin J-P, Petit L, Mitescu C 2001 {\it Physical Hydrodynamics} ed. Oxford
\bibitem{Konzuk2004} Konzuk J and Kueper H 2004 \textit{Water Resour. Res.} {\bf{40}} W02402
\bibitem{Meheust2000}M{\'e}heust Y and Schmittbuhl J 2000 \textit{Geophys. Res. Lett.} {\bf{27}} (18) 2989--92
\bibitem{Mourzenko1995}Mourzenko V, Thovert JF and Adler P 1995 \textit{J. Phys. II France} {\bf{5}} 465--82
\bibitem{Brown1995}Brown S, Stockman H, Reeves S 1995 \textit{Geophys. Res. Lett.} {\bf{22}} 2537--40
\bibitem{Gelhar1986}Gelhar L 1986 \textit{Water Resour. Res.} {\bf{22}} (1986) 135--45
\bibitem{Inoue2003}Inoue J and Sugita H 2003 \textit{Water Resour. Res.} {\bf{39}} 1202–-12
\bibitem{Zimmerman1991} Zimmerman R, Kumar S and Bodvarsson G
1991 \textit{Int. J. Rock Mech. Min. Sci.} {\bf{28}} (4) 325--31
\bibitem{Matsuki1999}Matsuki K, Lee J and Sakaguchi K 1999 \textit{Geotherm. Sci. Technol} {\bf{6}} 113--38
\bibitem{Brown1989}Brown S 1989 \textit{J. Geophys. Res.} {\bf{94}} 9429--38
\bibitem{Zimmerman1992}Zimmerman R, Chen D and Cook N 1992 \textit{J. Hydrol.} {\bf{139}} 79--96
\bibitem{Nolte1988}Pyrak-Nolte L, Cook N and Nolte D 1988 \textit{Geophys. Res. Lett.} {\bf{15}} 11 1247--50
\bibitem{Plouraboue2004}Plourabou{\'e} F, Geoffroy S and Prat M 2004 \textit{Phys. Fluids} {\bf{16}} (3) 615--24
\bibitem{Drazer2004}Drazer G, Auradou H, Koplik J and Hulin JP 2004 \textit{Phys. Rev. Lett.} {\bf{92}} 014501


\bibitem{Fardin2003}Fardin N 2003 {\it{The effect of scale on the
morphology, mechanics and transmissivity of single rock fracture}}, Dr
Thesis, Stockholm, Sweden
\bibitem{Moreno1988}Moreno L, Tsang Y, Tsang C, Hale V and Neretnieks I 1988 \textit{Water Resour. Res.} {\bf{24}} (12) 2033--48
\bibitem{Tsang1989} Tsang YW and Tsang CF 1989 \textit{Water Resour. Res.} {\bf{25}} (9) 2076--80
\bibitem{Bruderer2001}Bruderer C and Bernab{/'e} Y 2001 \textit{Water Resour. Res.} {\bf{37}} (4) 897--908
\bibitem{Gentier1997} Gentier S, Lamontagne E, Archambault G. and
Riss J 1997 \textit{Int. J. Rock Mech. Min. Sci. Geomech. Abstr.} {\bf{34}} 3--4
\bibitem{Auradou2001}Auradou H, Hulin JP and Roux S 2001 \textit{Phys. Rev. E} {\bf{63}} (6) 066306
\bibitem{Giacomini2008}Giacomini A, Buzzi O, Ferrero A, Migliazzaa M and Giania G 2008 \textit{Int. J. Rock Mech. \& Min. Sci.} {\bf{45}} (1) 47--58
\bibitem{Cardenas2007} Cardenas B, Slottke D, Ketcham R and Sharp J 2007 \textit{Geophys. Res. Lett.} {\bf{34}} L14404
\bibitem{Yiguang2008}Yiguang Y and Koplik J 2008 \textit{Phys. Rev. E} {\bf{77}} 036315
\bibitem{Skjetne1999}Skjetne E, Hansen A and Gudmundsson J 1999 \textit{J. Fluid Mech.} {\bf{383}} 1--28
\bibitem{Zimmerman2004}Zimmerman R, Al-Yaarubi A, Pain C and Grattoni C 2004 \textit{Int. J. Rock Mech. Min. Sci.} {\bf{41}} 3 Paper 1A 27
\bibitem{Mandelbrot1984}Mandelbrot B, Passoja D and Paullay A 1984 \textit{Nature} {\bf{308}} 721--22
\bibitem{Backers2003}Backers T, Fardin N, Dresena G and Stephanssona O 2003 \textit{Int. J. Rock
Mech. Min. Sci.} {\bf{40}} 425–33
\bibitem{Lee2003}Lee J, Kang J and Choe J 2003 \textit{Water Resour. Res.} \textit{39} (1) 1015
\bibitem{Neretnieks1982}Neretnieks I, Eriksen T and Tahtinen P 1982 \textit{Water Resour. Res.} \textit{18} 849--58
\bibitem{Keller1999}Keller A, Roberts P and Blunt M 1999 \textit{Water Resour. Res.} \textit{35} 55--63
\bibitem{Bear}Bear J 1972 \textit{Dynamics of Fluids in Porous Media} (New York: Elsevier) p~764
\bibitem{Detwiler2000}Detwiler R, Rajaram H and Glass R 2000 \textit{Water Resour. Res.} \textit{36} 1611--25
\bibitem{Park1997}Park C, Vandergraaf T, Drew D, Hahn PS 1997 \textit{J. Cont. Hydr.} {\bf{26}} 97--108
\bibitem{Ippolito1993}Ippolito I, Hinch E, Daccord G and Hulin JP 1993 \textit{Phys. Fluids A} {\bf{5}} (8) 1952--61
\bibitem{Boschan2008}Boschan A, Ippolito I, Chertcoff R, Auradou H, Talon L, Hulin JP 2008 \textit{Water Resour. Res.} {\bf{44}} W06420
\bibitem{Adler1999}Adler P, Thovert JF 1999 \textit{Fractures and Fracture Networks} Springer, Dordrecht, Netherland
\bibitem{Roux1998}Roux S, Plourabou{\'e} F and Hulin JP 1998 \textit{Transp. Porous Media} {\bf{32}} 97--116
\bibitem{Boschan2007}Boschan A, Auradou H, Ippolito I, Chertcoff R and Hulin JP 2007
\textit{Water Resour. Res.}  {\bf{43}} W03438
\bibitem{Boschan2009}Boschan A, Auradou H, Ippolito I, Chertcoff R and Hulin JP 2009 \textit{Water Resour. Res.} doi:10.1029/2008WR007461 in press
\bibitem{Auradou2006}Auradou H, Drazer G, Boschan A, Hulin JP and Koplik J 2006
\textit{Geothermics} {\bf{35}} 576--88
\bibitem{Aris1956}Aris R 1956 \textit{Proc. R. Soc. London, Ser. A} {\bf{235}} 67--77
\bibitem{Bodin2003}Bodin J, Delay F and de Marsily G 2003 \textit{Hydrogeol. J.} {\bf{11}} 418--33
\end{thebibliography}
\end{document}